\documentclass[prd,aps, preprint]{revtex4-1}
\usepackage{geometry}                % See geometry.pdf to learn the layout options. There are lots.
\usepackage{graphicx}
\usepackage{amssymb}
\usepackage{mathtools}
\usepackage{epstopdf}
\usepackage{verbatim}
\usepackage{color}
\usepackage{slashed}
\usepackage{bbold}
\usepackage{fullpage}
\usepackage{float}

\usepackage[title]{appendix}
\usepackage[colorlinks, citecolor=red, linkcolor=blue]{hyperref}

\newcommand{\beq}{\begin{equation}}
\newcommand{\eeq}{\end{equation}}
\newcommand{\bmat}{\begin{pmatrix}}
\newcommand{\emat}{\end{pmatrix}}
\newcommand{\bal}{\begin{align}}
\newcommand{\eal}{\end{align}}

\newcommand{\gnorm}{\sqrt{\frac{3}{5}}}
\newcommand{\MS}{\overline{MS}}
\newcommand{\DR}{\overline{DR}}

\DeclareMathOperator{\Tr}{Tr}

\DeclareMathOperator{\GeV}{GeV}

\begin{document}

%\title{ High-scale thresholds, $\Delta\lambda$ plots, \\and the viability of exact gauge coupling unification}

\title{Visualizing gauge unification with high-scale thresholds}

\author{Sebastian~A.~R.~Ellis}
%\email{sarellis@umich.edu}
\author{James~D.~Wells}
%\email{jwells@umich.edu}
%\affiliation{\it{Michigan Center for Theoretical Physics (MCTP),}\\ \it{Department of Physics, University of Michigan}, \\ \it{Ann Arbor, MI 48109, USA}}

\affiliation{\it{Michigan Center for Theoretical Physics (MCTP)}\\ \it{Department of Physics, University of Michigan, Ann Arbor, MI 48109, USA}}

\begin{abstract}
Tests of gauge coupling unification require knowledge of thresholds between the weak scale and the high scale of unification. If these scales are far separated, as is the case in most unification scenarios considered in the literature, the task can be factorized into IR and UV analyses. We advocate ``$\Delta\lambda$ plots" as an efficient IR analysis projected to the high scale. The data from these plots gives an immediate qualitative guide to the size of threshold corrections needed at the high scale (e.g., the indices of high-scale representations) and provides precise quantitative data needed to test the viability of hypothesized high-scale unification theories. Such an approach shows more clearly the reasonable prospects of non-supersymmetric grand unification in large rank groups, and also shows the low summed values of high-scale threshold corrections  required for supersymmetric unification. The latter may imply tuned cancellations of  high-scale thresholds in theories based on weak-scale supersymmetry. For that reason we view non-supersymmetric unification to be just as viable as supersymmetric unification when confining ourselves only to the question of reasonable high-scale threshold corrections needed for exact unification. We illustrate these features for a non-supersymmetric $SO(10)$ graund unified theory and a supersymmetric $SU(5)$ theory.

\end{abstract}
\maketitle

%\tableofcontents

\section{Introduction}

A common practice in the literature when contemplating gauge coupling unification is to settle upon a weak-scale theory, run the three gauge couplings up to the high scale, and look for theories where the three gauge couplings meet in one place. Famously, the Standard Model (SM) does not unify under that rubric but the Minimal Supersymmetric Standard Model (MSSM) does according to many \cite{Georgi:1974sy, Georgi:1974yf, Fritzsch:1974nn, Dimopoulos:1981zb, Dimopoulos:1981yj, Ibanez:1981yh, Ellis:1990zq}.  However, neither statement is true. The SM cannot be ruled out as the IR manifestation of a Grand Unified Theory (GUT), nor do the three gauge couplings meet at precisely one point in the MSSM. The key to understanding both claims is that high-scale thresholds are generically expected, which are the multiplets at the high scale that get masses by the same mechanism that breaks the GUT symmetry. High-scale threshold corrections kick the couplings further into exact unification if the underlying theory is gauge-coupling unified.

Despite continuing misstatements by a few at times, these facts have been well known by the experts for some time. What is not as widely appreciated is how generically possible unification is in theories without supersymmetry when expected high scale threshold corrections are contemplated, and how anomalously low the high-scale threshold corrections must be in supersymmetry to satisfy exact unification compared to generic expectations of a high-scale unified theory.  One of our main goals in this paper is to argue these two points by making apples to apples comparisons of the renormalized  gauge couplings at high scale and expected thresholds. 

The second main goal of this work is to transform all the information we have of the low-energy theory into useful data for testing the viability of a high-scale theory of unification.  The resulting output also should provide intuitive and immediate meaning to the unification seeker.  For example, RGE flow of $g_i$ couplings up to the high scale never achieve exact unification, even within supersymmetric theories. Indeed it should not if there are any high scale thresholds at all. The question then becomes whether the mismatch is too much for a viable unification theory to overcome. Simple data on $g_i$ at an (ambiguously defined) GUT scale is not enough to answer that question even approximately.   Further processing of the data is required.

In this article we advocate the answer to these requirements are plots of the scale-projected mismatch of couplings vs.\ the renormalization group running scale: ``$\Delta\lambda$ plots" . Definitions and details are below. Suffice it to say in the introduction here that these plots encapsulate all the needed information about the infrared thresholds of the theory, unambiguously show what high-scale threshold corrections need to accomplish to achieve exact unification of the couplings, and provide rapid intuition about the generic features that a unified theory must possess to have exact unification (e.g., the approximate size of representations needed).  All low-scale theories, including the SM and various forms of low-scale supersymmetry, need produce only one plot for researchers to use in testing viability of their high-scale unified theories.

The literature contains many examples of grand unified theories and their analyses, building on long-ago studies~\cite{Georgi:1974sy, Georgi:1974yf, Fritzsch:1974nn, Buras:1977yy, Dimopoulos:1981zb, Dimopoulos:1981yj, Ibanez:1981yh, Ellis:1990zq, Weinberg:1980wa, Hall:1980kf, Giunti:1991ta, Langacker:1992rq, Murayama:2001ur}. Our field continues to pursue grand unification in both supersymmetric contexts and not. Although weak scale supersymmetric grand unified theories are decidedly still viable~\cite{Anandakrishnan:2014nea}, many recent supersymmetry studies also consider the salient aspects of very massive superpartners~\cite{Acharya:2008zi, Hisano:2013cqa,Hisano:2013exa,Hall:2013eko,Hebecker:2014uaa,Wang:2015mea}. These theories require somewhat larger threshold corrections at the high scale, which as we will see may be a positive feature due to weak-scale supersymmetry's requirement of disquietingly small high-scale threshold corrections to achieve exact unification.
We demonstrate the utility of this method by applying it to two simple grand unified theories in the literature: The $SO(10)$ SM theory of Lavoura-Wolfenstein \cite{Lavoura:1993su} and the $\epsilon$-assisted $SU(5)$ supersymmetric theory of Tobe and Wells \cite{Tobe:2003yj}. Both of these theories are straighforwardly compatible with unification, despite the ``running of the couplings" not meeting exactly at any one scale. The $\Delta\lambda$ plots will be used to demonstrate the results graphically.

\section{Unification of couplings : preliminaries}

Let us begin by reviewing a few basic aspects of unification theory and effective field theory. When discussing the unification of couplings it is of great importance that we make unambiguous statements about what we know from the IR effective theory and what we can calculate from the UV in a particular GUT theory. It is possible to define a unified coupling $g_U$ in an infinite number of ways that are not physically meaningful. For example, we can define it to be the value of the couplings $g_1$ and $g_2$ at a scale $M_U$ where $g_1(M_U)=g_2(M_U)$ in some scheme. Or we can definite it to be when $g_2(M_U)=g_3(M_U)=g_U$, or the value of $(g_1+g_2+g_3)/3$ when $(g_1-g_2)^2+(g_1-g_3)^2+(g_2-g_3)^2$ is minimized, or an infinite number of other ways. Although one or more of these definitions can have some utility in some limited circumstances, in careful testing of theories for exact unification it is not useful to define a unified coupling $g_U$ by any procedure from the IR perspective.

Instead, $g_U$ can only be defined from the UV perspective where in the high-energy phase, or GUT phase, of the theory there is a single gauge coupling $g_U$ which is subject to defining boundary conditions to set its value at some scale, and which subsequently runs with scale. At some scale $M_*$ matching is made between the GUT unified theory with gauge coupling $g_U$ and the low-scale theory with gauge couplings $g_1$, $g_2$ and $g_3$. The matching at $M_*$ involves a lot of violence from threshold corrections, and $g_i(M_*)$ can all be quite different than $g_U$. For that reason there is little utility in trying to define a physically meaningful $g_U$ from the IR perspective. We shall therefore not rely on such artifices below.

%%%%%%%%%%%%%%%%%
\section{Analytic Definitions and Procedures}

Let us continue with some technical remarks on the calculations involved. The hypercharge, weak and strong couplings in the IR are the standard $g_1,~g_2,~g_3$, where $g_1$ has the appropriate GUT normalization. Their values change with scale according to the renormalization group equations (RGEs) of the IR effective theory. From the UV perspective, $g_U$ is defined by our choice of GUT theory. Depending on the choice of UV theory, there may be more than one scale involved depending on the splitting of the gauge boson and scalar masses. 
%We therefore define the scale at which unification occurs, $M_U$, as being the mass of the heaviest gauge boson in the theory. 

We use the two-loop RGEs for the evolution of the gauge coupling constants in the IR from the electroweak scale to the high scale. The equations are
\beq
 \frac{d g_i}{dt} = \beta^{(1)}_i + \beta^{(2)}_i = \frac{b_i g_i^3}{16\pi^2} + \frac{g_i^3}{(16\pi^2)^2}\left[ \sum_{j=1}^3 B_{ij} g_j^2 + \sum_{a=u,d,e} C_i^a \Tr\left( \boldsymbol{Y}_a^\dagger \boldsymbol{Y}_a\right)\right]
\eeq
where $b_i$, $B_{ij}$ and $C_i^a$ are group coefficients that can be calculated in the Standard Model and its extensions~\cite{Machacek:1983tz,Martin:1993zk}.

Rather than pick a particular unification scale, we choose a scale $\mu_* =10^{16}\GeV$ at which to evaluate various quantities. We select this scale since it is closely related to the constraints on the masses of the vector bosons associated with proton decay. We know that at scales near the unification scale, the IR gauge couplings $g_i(\mu_*)$ are related to the unification coupling $g_U(\mu_*)$ by the following relation at one-loop \cite{Weinberg:1980wa, Hall:1980kf}:
\beq
\label{BCunif.EQ}
\left(\frac{1}{g_i^2(\mu_*)}\right)_{\MS} = \left(\frac{1}{g_U^2(\mu_*)}\right)_{\MS} - \left(\frac{\lambda_i(\mu_*)}{48\pi^2}\right)_{\MS}
\eeq
where $\lambda_i(\mu_*)$ are the threshold corrections, computed in the $\overline{MS}$ scheme, to each gauge coupling at the scale $\mu^*$. In general, when masses in an irreducible block are identical, $\lambda_i(\mu)$ can be defined as \cite{Hall:1980kf}
%\beq
%\label{Thresh.EQ}
%\left(\lambda_i(\mu)\right)_{\MS}= \Bigg\{ \Tr(t_{i,V}^2) -21 \Tr\left( t_{i,V}^2 \ln\frac{M_V}{\mu}\right) + \Tr\left( t_{i,S}^2 \Lambda \ln\frac{M_S}{\mu} \right) + 8\Tr\left( t_{i,F}^2 \ln\frac{M_F}{\mu}\right) \Bigg\}
%\eeq
%where $M_X,~X=V,~S,~F$ is the mass of the superheavy gauge boson, scalar and fermion respectively, with associated generators $t_X$. The projection operator $\Lambda$ is defined such that $\Lambda\phi$ are the physical heavy scalars, and $(1-\Lambda)\phi$ are the Goldstone bosons. The trace is performed over the matrix of heavy particle masses $M_X$. Because the masses in an irreducible block are not necessarily the same, we leave the implicit summation notation, rather than as a Casimir multiplying a log. In the case where masses are identical, we may write Eq. (\ref{Thresh.EQ}) above as
\begin{align}
\left(\lambda_i(\mu)\right)_{\MS}=   l^{V_n}_{i} -21\,  l^{V_n}_{i} \ln\frac{M_{V_n}}{\mu} + l^{S_n}_{i} \ln\frac{M_{S_n}}{\mu}  + 8\, l^{F_n}_{i} \ln\frac{M_{F_n}}{\mu} 
\end{align}
 where there is an implicit sum over the $n$ different superheavy particles of a given type. 
 %Additionally we have removed the $\Lambda$ projection operator, but
 It should be understood that only physical scalars contribute. The $l^X_i$ are the weighted Dynkin indices relative to the SM gauge group $i$.
This computation of $\lambda_i(\mu)$ is understood to be accurate only in the region near the scale of unification. The threshold corrections can therefore be determined in the GUT theory of choice.

In the IR, we may use Eq. (\ref{BCunif.EQ}) above and define the following relations that are independent of the unification coupling $g_U(\mu_*)$
\begin{align}
\label{lij.EQ}
\left(\frac{\Delta\lambda_{ij}(\mu_*)}{48\pi^2}\right)_{\MS,~\DR} \equiv \left(\frac{1}{g_i^2(\mu_*)}-\frac{1}{g_j^2(\mu_*)}\right)_{\MS,~\DR} = \left(\frac{\lambda_j(\mu_*) - \lambda_i(\mu_*)}{48\pi^2}\right)_{\MS,~\DR}
\end{align}
for $i,j=1~,2,~3,~i\neq j$.
Any two $\Delta\lambda_{ij}$ then specify all the threshold corrections up to a constant factor. The subscripts $\MS$ and $\DR$ indicate that the threshold corrections and gauge couplings need to be computed in the appropriate renormalization scheme depending on whether one is dealing with a SUSY theory ($\DR$) or not ($\MS$).

From the IR, we only know how to compute $g_i(\mu)$ and run up to some scale $\mu_*$. We may then use Eq. (\ref{lij.EQ}) to calculate $\Delta\lambda_{ij}$ as a function of $\mu$ without requiring knowledge of the UV theory. We may then assume that the UV has some GUT theory description, which would allow us to compute the threshold corrections $\lambda_i(\mu_*)$ and their difference, $\Delta\lambda_{ij}(\mu_*)$ given the spectrum of superheavy particles. If the $\Delta\lambda_{ij}(\mu_*)$ obtained from the IR were to match that obtained from the UV GUT theory for a particular set of GUT masses, unification is possibly achieved. There is an ambiguity due to a shift symmetry since the differences matching may not account for a constant term. Therefore matching the UV and IR calculations of $\Delta\lambda_{ij}$ specifies 
\beq
\frac{1}{g_U^2(\mu_*)} + S~~~{\rm and}~~~ \frac{\lambda_i(\mu_*)}{48\pi^2}+S
\eeq
where $S$ is some constant shift. Thus Eq. (\ref{BCunif.EQ}) is satisfied, but with some ambiguity from the IR perspective left over as to the unified coupling constant and the size of the threshold corrections. Of course, from the UV perspective, both are known.

%%%%%%%%%%%%%%%%%%%%%%%%%%%%%%%
\section{Numerical Procedure}

We compute the running of the gauge couplings in three different cases: (1)  the Standard Model (SM); (2) a generic low-scale CMSSM-like SUSY model with weakly (strongly) coupled sparticles at 1 TeV (3 TeV); and, (3) a split supersymmetry-like model with bino and higgsinos at 1 TeV, winos at 3 TeV, gluinos at 7 TeV and scalar superpartners at 1 PeV.  These three cases are all motivated by theory and current experimental constraints from the LHC.

We perform the calculation of the RGE running at two-loop order numerically, with the one-loop coefficients changing depending on the matter content.  For the SM, we use the two-loop beta functions derived in \cite{Machacek:1983tz}. 
For the supersymmetric cases, we use the two-loop beta functions derived in~\cite{Martin:1993zk} with the appropriate shifts in the one-loop coefficients as we pass through sparticle thresholds. 

\begin{figure}[t]
\includegraphics[scale=0.3]{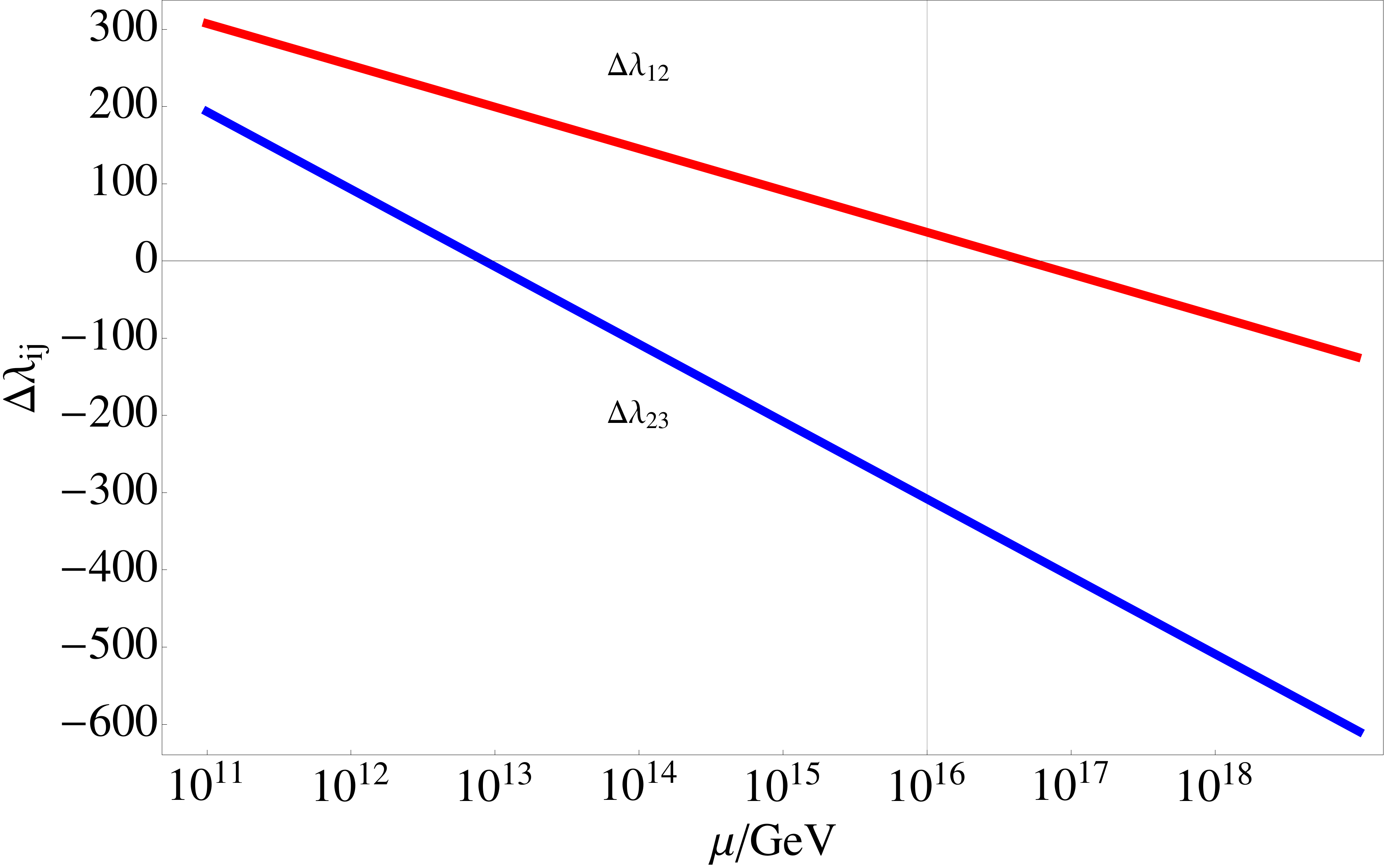}
\caption{Plot of $\Delta\lambda_{23}(\mu)$ (red) and $\Delta\lambda_{12}(\mu)$ (blue) for the Standard Model as a function of scale $\mu$.}
\label{fig:SM-Delta-lambda}
\end{figure}

Upon applying these spectra to the above formalism we are now able to compute the $\Delta\lambda$ plot of $\Delta\lambda_{13}$ and $\Delta\lambda_{23}$ values from the IR for all three cases as a function of renormalization scale $\mu$. The SM case is provided in Fig.~\ref{fig:SM-Delta-lambda}. This plot is the only data needed when one wishes to check SM compatibility with a favorite unification theory. We will do that later in an $SO(10)$ example.

\begin{figure}[t]
\includegraphics[scale=0.3]{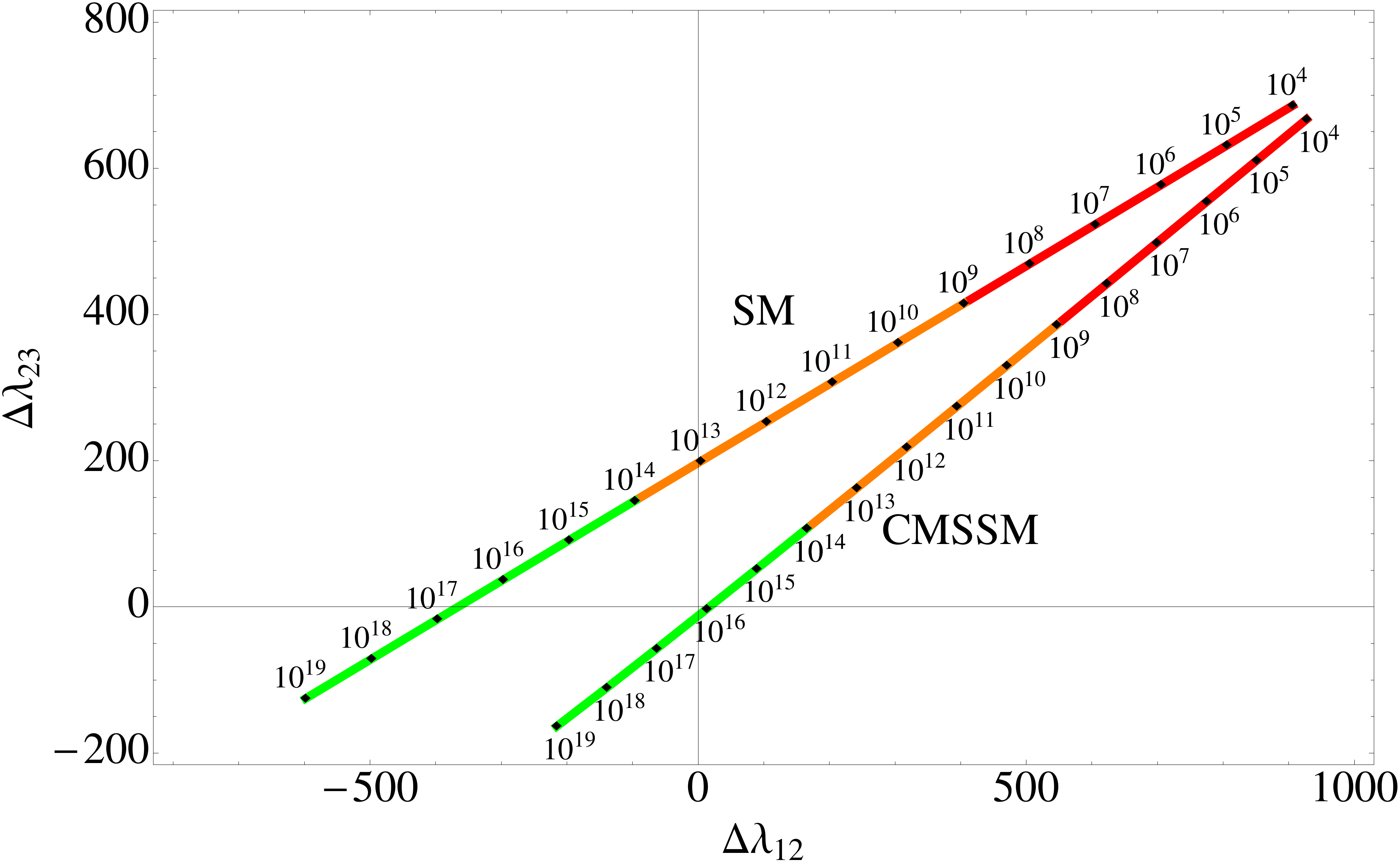}
\caption{This key visualization plot shows $\Delta\lambda_{23}(\mu)$ as a function of $\Delta\lambda_{12}(\mu)$ for the Standard Model and a CMSSM-like SUSY model. Labels on the line indicate the scale $\mu$. Green regions indicate that a unification scale around those values is moderately safe from constraints. Orange indicates relatively unsafe, Red indicates very unsafe.}
\label{A12vsA23.FIG}
\end{figure}

Each supersymmetric theory with a well-defined spectrum of superpartners has its own unique $\Delta\lambda$ plot. In our two supersymmetric cases (2) and (3) discussed above, it is straightforward to make a plot of $\Delta\lambda_{13}$ and $\Delta\lambda_{23}$ as a function of $\mu$ for each of them as we did for the SM case in Fig.~\ref{fig:SM-Delta-lambda}. However, there is an even simpler representation of the same information that we wish to present. One can plot the correlated values of $\Delta\lambda_{23}$ vs.\ $\Delta\lambda_{12}$ parametrized by the renormalization scale $\mu$, where the values of $\mu$ are labeled on the line. This is done for the SM and the CMSSM-like SUSY model (case 2 above) in Fig.~\ref{A12vsA23.FIG}. 

To better aid the vision and intuition of these plots we color code the line into red, orange and green as we cross various thresholds of $\mu$.  We suggestively call green ``safe" to follow intuitions of simple grand unified groups that thresholds above $10^{14}\GeV$ (at least) are needed to protect the proton from decaying too quickly.  

We note in Fig.~\ref{A12vsA23.FIG} that the supersymmetric line crosses very close to zero for $\mu\simeq 10^{16}\GeV$ which is illustrating the famous case for supersymmetric unification. 
%By passing through the origin it appears that there is no mismatch between the couplings at this high scale and unification occurs. 
The SM line strays far from the origin of the plot and illustrates the famous case against SM unification. However, what precisely does it mean to ``stray far" from the origin? How far is too far?  The answer to these questions starts by acknowledging that exact unification at the high scale, and a high-scale theory, both require analyzing the high-scale threshold corrections that are generically expected. A line for a theory in the $\Delta\lambda$ plot, such as the SM line in Fig.~\ref{A12vsA23.FIG}, is ``too far" away from the origin if we cannot imagine threshold corrections at the high scale shifting the couplings enough to bring it back to the origin.  

We will see below that in the case of supersymmetry, there is never a problem in this regard. In fact, the $\Delta\lambda$'s are arguably too small and threshold corrections have to either not be present for some reason or must have tuned cancellations at the high scale for exact unification to occur. In the case of the SM the corrections are large, and the index of the representations at the high scale must be comparable to the $\Delta\lambda$ values (up to multiplicative logarithms) of up to several hundred. However, the index of representations of grand unified theories based on $SO(10)$ are often in the three digits, such as the {\bf 126} representation with index 35 and the {\bf 210} representation with index 56~\cite{Slansky:1981yr}. Indeed, these representations play a key role in our first example of the next section: Lavoura and Wolfenstein's non-supersymmetric $SO(10)$ theory~\cite{Lavoura:1993su}.

%%%%%%%%%%%%%%%%%%%%%%%%%%%%
\section{Exact Unification Examples}

%The unification of the strong, weak and electromagnetic (EM) interactions into a grand unified theory (GUT) based on a simple group $G$ at very large energies is an attractive proposition. Many such groups have been suggested, such as $SU(5)$  or $SO(10)$, both in the context of the Standard Model \cite{Georgi:1974sy, Fritzsch:1974nn} and in the case of Supersymmetry (SUSY) \cite{Dimopoulos:1981zb}.  It is within the grand unified theory context that the techniques described in this paper are most useful.
 
Having described the theoretical framework we use, we give below two examples of grand unified theories that illustrate the viability of SM and supersymmetric GUTs, demonstrate the utility of the $\Delta\lambda$ plots in the search for a precise spectrum that yields exact unification of the couplings. The two examples are Lavoura and Wolfenstein's non-supersymmetric $SO(10)$ theory~\cite{Lavoura:1993su} and Tobe-Wells supersymmetric $SU(5)$ theory~\cite{Tobe:2003yj}.

\subsection{Lavoura-Wolfenstein SO(10) and the Standard Model}

The $SO(10)$ GUT of Lavoura and Wolfenstein \cite{Lavoura:1993su} has a  Higgs structure that consists of  $\{\boldsymbol{10}$ and $\boldsymbol{126}\}$ representations. There is also a $\boldsymbol{210}$ which contains heavy scalars that do not condense.

The Lavoura-Wolfenstein Spectrum is shown in table \ref{LWspectrum.TAB} below. We label the gauge bosons according to their SM gauge group representations to make clear that the gauge bosons leading to proton decay are all at the common mass $M_V$, and those not leading to proton decay are all at a different mass $M_R$. The mass $M_R$ is defined to be $M_R \approx v_R$, where $v_R$ is the VEV of the $(1,3,10)$ of the $\boldsymbol{126}$. We label the scalars according to their $SU(2)_L \otimes SU(2)_R \otimes SU(4)$ representations. This is done because the simplifying assumption is made that all the SM representations of scalars in a given $SU(2)_L \otimes SU(2)_R \otimes SU(4)$ representation will have the same mass. The $(2,2,1)$ component of the $\boldsymbol{10}$ contains the SM Higgs and therefore is not listed, as it will not contribute to the threshold corrections at the unification scale.
\begin{table}
\begin{tabular}{|c | c | c| c | c | c|}
\hline
\multicolumn{3}{|c|}{\textbf{Gauge Bosons}} & \multicolumn{3}{|c|}{\textbf{Scalars}} \\
\hline
$SO(10)$ & $SU(2) \otimes SU(3) [U(1)_Y]$ & Mass & $SO(10)$ & $  SU(2)_L \otimes SU(2)_R \otimes SU(4)$ & Mass \\
\hline
$\boldsymbol{45}$ & $(1,1)[0]$ & $M_R$ & $\boldsymbol{210}$& $(1,1,1)$ & N/A \\
$\boldsymbol{45}$ & $(1,1)[\sqrt{\frac{3}{5}}]$ & $M_R$ & $\boldsymbol{210}$ & $(2,2,6)$ & Goldstone\\
$\boldsymbol{45}$ & $(1,1)[-\sqrt{\frac{3}{5}}]$ & $M_R$ & $\boldsymbol{210}$ &$(1,1,15)$ & $M_1$\\
$\boldsymbol{45}$ & $(1,3)[\frac{2}{3}\sqrt{\frac{3}{5}}]$ & $M_R$ & $\boldsymbol{210}$ & $(2,2,10)$& $M_1$\\
$\boldsymbol{45}$ & $(1,\overline{3})[-\frac{2}{3}\sqrt{\frac{3}{5}}]$ & $M_R$ & $\boldsymbol{210}$ &$(2,2,\overline{10})$ & $M_1$\\
$\boldsymbol{45}$ & $(2,3)[\frac{1}{6}\sqrt{\frac{3}{5}}]$ & $M_V$ & $\boldsymbol{210}$& $(1,3,15)$& $M_4$\\
$\boldsymbol{45}$ & $(2,\overline{3})[-\frac{1}{6}\sqrt{\frac{3}{5}}]$ & $M_V$ &$\boldsymbol{210}$ &$(3,1,15)$ & $M_5$\\ \cline{4-6}
$\boldsymbol{45}$ & $(2,3)[-\frac{5}{6}\sqrt{\frac{3}{5}}]$ & $M_V$ & $\boldsymbol{126}$& $(1,1,6)$ & $M_1$\\
$\boldsymbol{45}$ & $(2,\overline{3})[\frac{5}{6}\sqrt{\frac{3}{5}}]$ & $M_V$ &$\boldsymbol{126}$ & $(2,2,15)$ &$M_1$\\
&&& $\boldsymbol{126}$ & $(1,3,10)$ & $M_2$ \\
&&& $\boldsymbol{126}$ & $(3,1,\overline{10})$ & $M_3$ \\
\hline
\end{tabular}
\caption{Table showing the spectrum of superheavy particles contributing to the threshold corrections in the Lavoura-Wolfenstein SO(10) GUT, with their various masses.}
\label{LWspectrum.TAB}
\end{table}
The decomposition into the various SM representations for a given $SU(2)_L \otimes SU(2)_R \otimes SU(4)$ can be done. For example, the 
$(1,1,15)$ of the SO(10) $\boldsymbol{210}$ yields under $(SU(2),SU(3))_{U(1)_Y}$ the charges
\beq
%(1,1)[0],~(1,3)\left[\,\frac{2}{3}\gnorm\,\right],~(1,\overline{3})\left[\,-\frac{2}{3}\gnorm\,\right],~(1,8)[0]
(1,1)_0,~~(1,3)_{Q'},~~(1,\overline{3})_{-Q'},~~{\rm and}~~(1,8)_0,~~~{\rm where}~~Q'=\frac{2}{3}\gnorm.
\eeq 
Making the decomposition explicit is unnecessary for the purposes of defining the spectrum of masses, but it must be done in order to compute the contributions that each state will make to the threshold corrections.

The threshold corrections for this particular GUT are obtained by applying the boundary condition equation for the threshold corrections (Eq. (\ref{BCunif.EQ})) for each of the vector bosons and scalars that has a mass near the unification scale and summing over all heavy fields. Each heavy boson contributes
\beq
\left(\lambda_i^{V_n}\right)_{\overline{MS}} = l_i^{V_n} \left( 1 + 21 \ln \frac{\mu_*}{M_V}\right) 
\eeq
where $l_i^{V_n}$ is the Dynkin index of the $n$-th vector boson relative to the SM group labelled by $i$, multiplied by their dimensions relative to the other SM gauge groups. Each scalar contributes
\beq
 \left(\lambda_i^{S_n}\right)_{\MS} = - l_i^{S_n} \ln \frac{\mu_*}{M_S}
\eeq
with the same labels as before, with the Dynkin index for the scalar.
Given the content of the Lavoura-Wolfenstein SO(10) GUT, we obtain
\begin{align}
\lambda^V_1(\mu_*) &= 8 + \frac{294}{5} \log\frac{\mu_*}{M_R} + \frac{546}{5}\log\frac{\mu_*}{M_V}\\
\lambda^V_2(\mu_*) &= 6 + 126\log\frac{\mu_*}{M_V}\\
\lambda^V_3(\mu_*) &= 5 +21\log\frac{\mu_*}{M_R} + + 84\log\frac{\mu_*}{M_V}
\end{align}
for the contributions from vector bosons, and
\begin{align}
\lambda^S_1(\mu_*) &= -\frac{274}{5}\log\frac{\mu_*}{M_1} -  \frac{142}{5}\log\frac{\mu_*}{M_2}-  \frac{36}{5}\log\frac{\mu_*}{M_3} -  \frac{114}{5}\log\frac{\mu_*}{M_4}   \\
\lambda^S_2(\mu_*) &= -50\log\frac{\mu_*}{M_1} -40\log\frac{\mu_*}{M_3} -30\log\frac{\mu_*}{M_5}\\
\lambda^S_3(\mu_*) &= -62\log\frac{\mu_*}{M_1}-17\log\frac{\mu_*}{M_2}-18\log\frac{\mu_*}{M_3}-12\log\frac{\mu_*}{M_4}-12\log\frac{\mu_*}{M_5}
\end{align}
for the contributions from the scalars.

\begin{figure}[t]
\includegraphics[scale=0.3]{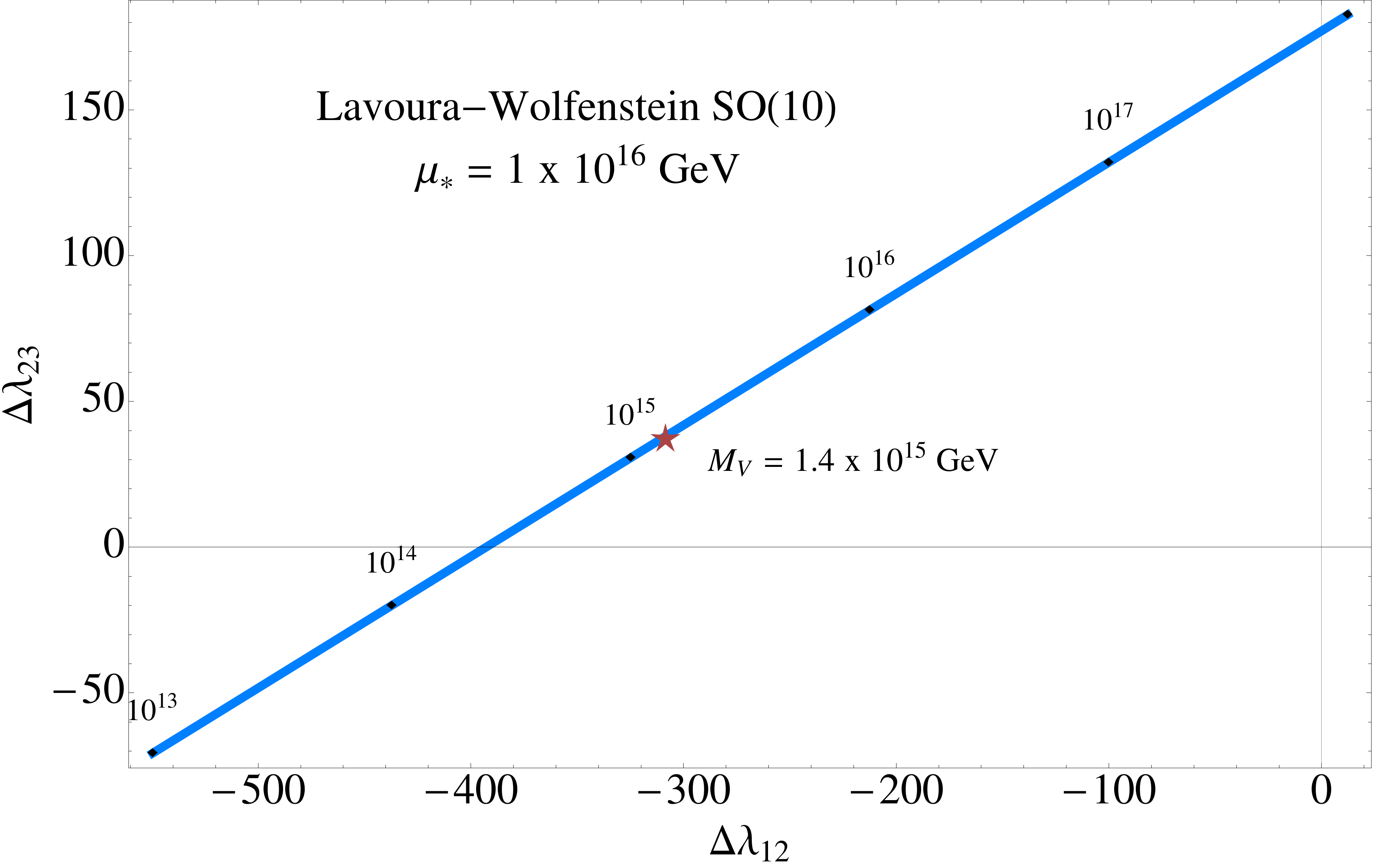}
\caption{Plot of $\Delta\lambda_{23}(\mu)$ as a function of $\Delta\lambda_{12}(\mu)$. Shown is the Lavoura-Wolfenstein SO(10)  (blue) with $M_V/M_R=20$, $M_V/M_1=3$, $M_V/M_2=7$, $M_V/M_3=8$, $M_V/M_4=10$ and $M_V/M_5=14$, with $M_V$ varying between $10^{13}$ and $10^{18}$. The star corresponds to the required values of $\Delta\lambda_{12}(\mu_*)$ and $\Delta\lambda_{23}(\mu_*)$ in the SM. We find that $M_V = 1.4 \times 10^{15}$ gives the desired $\Delta\lambda_{12}(\mu_*)$ and $\Delta\lambda_{23}(\mu_*)$ in the Lavoura-Wolfenstein SO(10) for the given mass ratios.}
\label{A12vsA23LavWolf.FIG}
\end{figure}

We consider a particular set of high-scale mass ratios, chosen to ensure intersection with the SM $\Delta\lambda_{ij}(\mu_*)$
\beq
\frac{M_V}{M_R}=20,~ \frac{M_V}{M_1}=3,~ \frac{M_V}{M_2}=7,~ \frac{M_V}{M_3}=8,~ \frac{M_V}{M_4}=10,~ \frac{M_V}{M_5} = 14.
\eeq
This enables us to evaluate $\Delta\lambda_{ij}(\mu_*)$ in the Lavoura-Wolfenstein GUT. Since we are interested in what happens if we modify the GUT particle masses, we vary $M_V$, keeping the ratios fixed. The resultant plot is shown in Fig. \ref{A12vsA23LavWolf.FIG}. The point of intersection with the SM $\Delta\lambda_{ij}(\mu_*)$  therefore fixes $M_V$ in the Lavoura-Wolfenstein SO(10). 

This example has now demonstrated how to process IR data in the form of the $\Delta\lambda$ plot of Fig.~\ref{A12vsA23.FIG}. Upon choosing a renormalization scale $\mu^*$ there is a single point in the  $\Delta\lambda_{23}-\Delta\lambda_{12}$ plane that is compatible with exact unification, and it  is required that high-scale thresholds must give those values. In Fig.~\ref{A12vsA23LavWolf.FIG} we show that indeed those values can be achieved for the spectrum specified in table~\ref{LWspectrum.TAB}, and exact unification therefore is shown to be viable.

%%%%%%%%%%%%%%%%%%%%%%%
\subsection{Tobe-Wells Supersymmetric SU(5)}

We also compare with a SUSY model, taking as an example an SU(5) GUT described by Tobe and Wells \cite{Tobe:2003yj}. The Higgs structure of this GUT consist of $\{ \boldsymbol{24_H}, \boldsymbol{5_H},\boldsymbol{\overline{5}_H}\}$, and the gauge representation is a $\boldsymbol{24}$. The spectrum of the superheavy particles in this GUT is shown in Table \ref{TWspectrum.TAB}.

\begin{table}
\begin{tabular}{|c | c |c|c | c | c|}
\hline
\multicolumn{3}{|c|}{\textbf{Gauge Bosons}} & \multicolumn{3}{|c|}{\textbf{Scalars}} \\
\hline
$SU(5)$ & $SU(2) \otimes SU(3) [U(1)_Y]$ & Mass & $SU(5)$ & $SU(2) \otimes SU(3) [U(1)_Y]$ & Mass \\
\hline
$\boldsymbol{24}$ & $(2,3)[-\frac{5}{6}\gnorm]$ & $M_V$ & $\boldsymbol{24}_H$ & All & $M_\Sigma$\\
$\boldsymbol{24}$ & $(2,\overline{3})[\frac{5}{6}\gnorm]$ & $M_V$ &$\boldsymbol{5}_H + \boldsymbol{\overline{5}}_H$ &$(1,3)[-\frac{1}{3}\gnorm] + (1,\overline{3})[\frac{1}{3}\gnorm]$& $M_{H_c}$\\
\hline
\end{tabular}
\caption{Table showing the spectrum of superheavy particles contributing to the threshold corrections in the Tobe-Wells SU(5) GUT, with their various masses.}
\label{TWspectrum.TAB}

\end{table}

In this GUT there is an additional non-renormalizable operator connecting the adjoint Higgs representation to the gauge fields. The operator in question arises from the gauge-kinetic function of minimal SU(5) written as
\beq
\int d^2\theta \left[\frac{S}{8M_{Pl}} \mathcal{W}\mathcal{W} + \frac{y\Sigma}{M_{Pl}}\mathcal{W}\mathcal{W}\right]
\eeq
where $\Sigma = \boldsymbol{24_H}$. The second term gives rise to corrections to the gauge couplings because the adjoint Higgs must acquire a vacuum expectation value
\beq
\langle \Sigma \rangle = v_\Sigma~ \text{diag}\left( \frac{2}{3},\frac{2}{3},\frac{2}{3},-1,-1\right)
\eeq
to break SU(5) to the SM gauge group at the GUT scale. The masses of the X and Y bosons are related to the vev of $\Sigma$ by
\beq
M_{X,Y}^2 = \frac{25}{18}g_U^2 v_\Sigma^2.
\eeq
%We keep our definition that the unification scale $M_U$ is the mass of the heaviest gauge boson, such that $M_U\equiv M_V = M_{X,Y}$.

The relationship between the $g_i$ and $g_U$ couplings is altered by a term depending on $\epsilon/ 48\pi^2 \equiv  8yv_\Sigma/M_{Pl} $
\begin{align}
\left(\frac{1}{g_i^2(\mu_*)}\right)_{\DR} = \left(\frac{1}{g_U^2(\mu_*)}\right)_{\DR} - \left(\frac{\lambda_i(\mu_*) - c_i \epsilon}{48\pi^2}\right)_{\DR}
\end{align}
where $c_i = \{-2/3,-1,~2/3\}$. This allows us to define a further contribution to $\Delta\lambda_{ij}$
\begin{align}
\Delta\lambda^\epsilon_{ij} = c_i \epsilon - c_j \epsilon
\end{align}
which can be included along with the contributions from the threshold corrections. 

Threshold contributions to $g_i(\mu^*)$  are given in the $\overline{DR}$ scheme by
\beq
\left(\lambda^{Heavy}_i\right)_{\DR} = 6\cdot l_{Heavy, i} \ln\frac{\mu_*}{M_{Heavy}} 
\eeq
where $l_{Heavy,i}$ is the weighted Dynkin index of a heavy particle of mass $M_{heavy}$ for the $i$-th SM gauge group (the factor of 6 is to obtain the appropriate normalization). We may then write the contributions from the various heavy particles that are not at the unification scale. For the heavy vector bosons, the contributions are
\beq
\lambda^{V}_1(\mu_*) = 60\log\frac{\mu_*}{M_V},~~~
\lambda^{V}_2(\mu_*) = 36\log\frac{\mu_*}{M_V},~~~{\rm and}~~~
\lambda^{V}_3(\mu_*) = 24\log\frac{\mu_*}{M_V}.
\eeq
Contributions from the heavy colored Higgs $H^c$  are
\beq
\lambda^{H_c}_1(\mu_*) = -\frac{12}{5}\log\frac{\mu_*}{M_{H_c}},~~~
\lambda^{H_c}_2(\mu_*) = 0,~~~{\rm and}~~~
\lambda^{H_c}_3(\mu_*) = -6\log\frac{\mu_*}{M_{H_c}},
\eeq
and those from $\Sigma$ are
\beq
\lambda^{\Sigma}_1(\mu_*) = 0,~~~
\lambda^{\Sigma}_2(\mu_*) = -12\log\frac{\mu_*}{M_\Sigma},~~~{\rm and}~~~
\lambda^{\Sigma}_3(\mu_*) = -18\log\frac{\mu_*}{M_\Sigma}.
\eeq

\begin{figure}[t]
\includegraphics[scale=0.3]{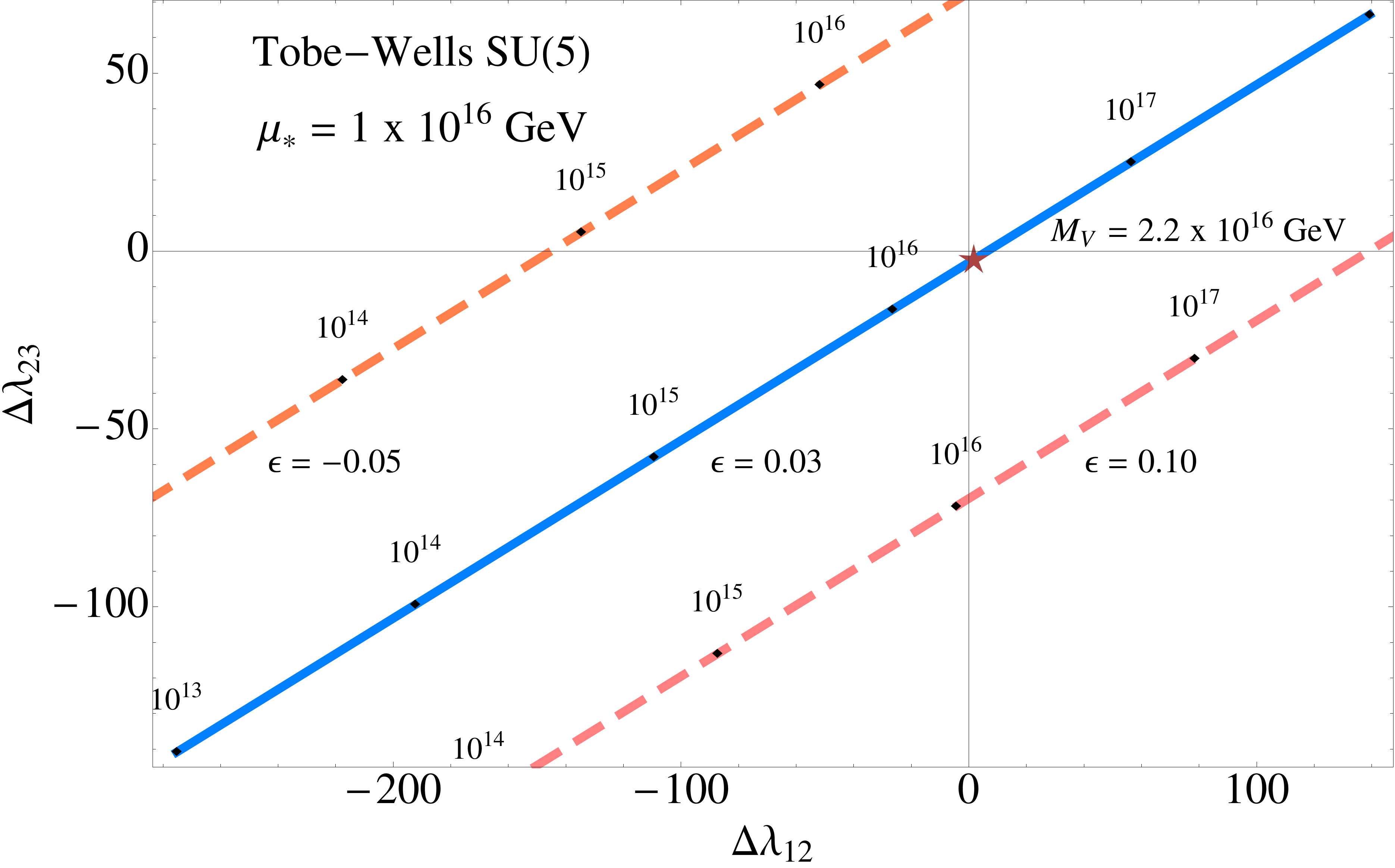}
\caption{Plot of $\Delta\lambda_{23}(\mu_*)$ as a function of $\Delta\lambda_{12}(\mu_*)$ for the CMSSM-like SUSY model. The value of $M_{H_c}$ was fixed to be $3.3\times10^{17}$, both to ensure coincidence with the CMSSM point, and to ensure avoidance of constraints. Three curves of different $\epsilon$ values show the effect of varying that parameter. The ratio $M_{\Sigma} / M_{H_c}$ was fixed at 0.1, and then $M_V$ was varied. We find that $M_V=2.2\times10^{16}$ in the Tobe-Wells SU(5) GUT yielded matching of $\Delta\lambda_{ij}(\mu_*)$ to the SM. }
\label{A12vsA23TobeWells.FIG}
\end{figure}

The plot for the CMSSM-like SUSY model shown in Fig. \ref{A12vsA23TobeWells.FIG} was made for fixed values of $M_{H_c}=3.3\times10^{17}$, $M_\Sigma = 0.1 \times M_{H_c}$ and $\epsilon = 0.03$. Then by varying $M_V$ we found the point of intersection of the Tobe-Wells SU(5) $\Delta\lambda_{ij}(\mu_*)$ with the $\Delta\lambda_{ij}(\mu_*)$ from the CMSSM-like IR theory.

\begin{figure}[t]
\includegraphics[scale=0.3]{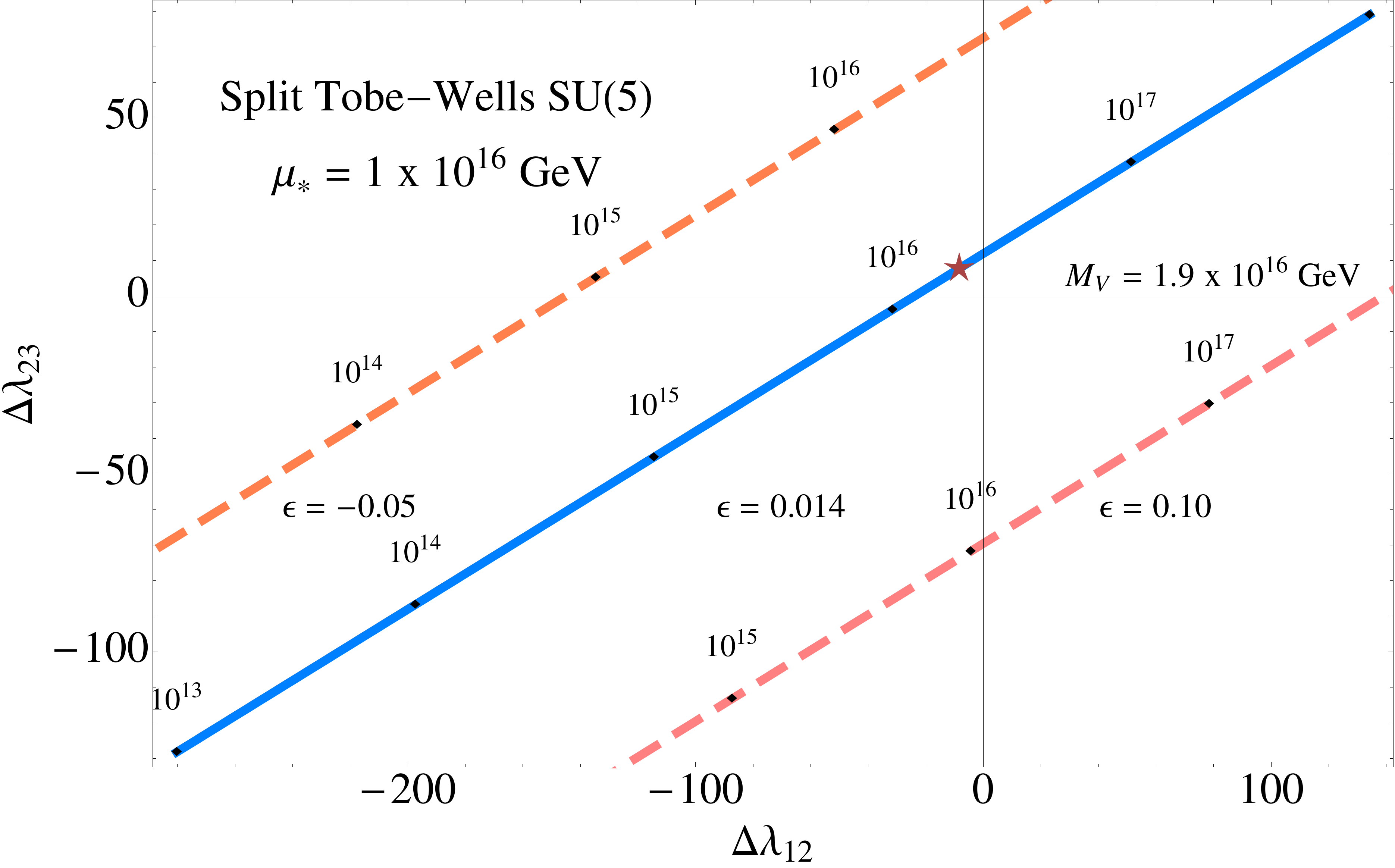}
\caption{Plot of $\Delta\lambda_{23}(\mu_*)$ as a function of $\Delta\lambda_{12}(\mu_*)$ for the Split-SUSY-like model. The value of $M_{H_c}$ was fixed to be $3.3\times10^{17}$, both to ensure coincidence with the Split-SUSY point, and to ensure avoidance of constraints. Three curves of different $\epsilon$ values show the effect of varying that parameter. The ratio $M_{\Sigma} / M_{H_c}$ was fixed at 0.1, and then $M_V$ was varied. We find that $M_V=1.9\times10^{16}$ in the Tobe-Wells SU(5) GUT yielded matching of $\Delta\lambda_{ij}(\mu_*)$ to the SM. }
\label{A12vsA23TobeWellsSplit.FIG}
\end{figure}

The plot for the  Split-SUSY-like model shown in Fig. \ref{A12vsA23TobeWellsSplit.FIG} was made for fixed values of $M_{H_c}=3.3\times10^{17}$, $M_\Sigma = 0.1 \times M_{H_c}$ and $\epsilon = 0.014$. Then by varying $M_V$ we found the point of intersection of the Tobe-Wells SU(5) $\Delta\lambda_{ij}(\mu_*)$ with the $\Delta\lambda_{ij}(\mu_*)$ from the Split-SUSY-like IR theory. This example again shows how the factorization of the IR data and the UV GUT data can be compared through $\Delta\lambda$ plot visualizations to establish viable exact unification of the gauge couplings.

%%%%%%%%%%%%%%%%%%%%
\section{Conclusions}

In this article we have reviewed the technical procedures for determining if a theory is compatible with exact unification of the gauge couplings. For theories with a large desert between the weak scale and the high scale where unification occurs, the problem conveniently factorizes into an analysis of the low-scale theory and the high-scale theory. 

We have demonstrated that the data needed from the low-scale theory to make this assessment is encapsulated well by $\Delta\lambda_{23}$ vs.\ $\Delta\lambda_{12}$ plots parametrized by the renormalization running scale $\mu$. We have constructed these plots for three different low-scale theories: the SM, a CMSSM-like supersymmetric theory, and a split supersymmetry theory. The results give us an immediate and intuitive understanding for the scale of gauge coupling unification and the size of corrections (i.e., the indices of high-scale representations) needed to achieve exact unification. They also provide all the information needed technically to perform a careful check of unification. 

We have illustrated this approach by matching the data from the SM $\Delta\lambda$ plot to the threshold corrections of the Lavoura-Wolfenstein non-supersymmetric $SO(10)$ theory. It is an example that demonstrates a general result, which is that non-supersymmetric gauge coupling unification is indeed possible without unexpectedly large threshold corrections in grand unified theories based on high-rank gauge group with large representations, such as $SO(10)$.  

We also have illustrated the approach by finding a spectrum in the Tobe-Wells $SU(5)$ theory that matches the needed threshold corrections implied by the SM $\Delta\lambda$ plot. This example illustrates the general point that supersymmetric unification requires either very small high-scale threshold corrections, or a partial cancellation of the threshold corrections to achieve exact unification. In the Tobe-Wells case, that cancellation is aided by a non-renormalizable coupling of the Higgs to the gauge kinetic function.  

We end by pointing out that any plots of the gauge couplings $g_i$ or $1/g_i$ or even $1/\alpha_i$ are of little value for deciding if a theory is favorable to gauge coupling unification. The $\Delta\lambda$ plot parametrized by the renormalization running scale $\mu$, which can be made for any well defined theory in the IR, is a significantly better way to collect and visualize the necessary data from the low-scale theory to apply to question of high-scale unification. Qualitative understanding of what is required of high-scale thresholds and technical data needed to make the assessment are contained within the $\Delta\lambda$ plot. We also believe that the physically meaningful $\Delta\lambda$ plots also show that when it comes to expected high-scale threshold corrections, unification within a non-supersymmetric theory is just as viable as within a supersymmetric theory. That is to say, unification is attractive in both approaches, and additional considerations are necessary to draw preferences.

\noindent
{\it Acknowledgements:} The work of JDW is supported in part by DoE grant DE-SC0011719 and the work of SARE is supported in part by DoE grant DE-SC0007859.

\end{document}